\documentclass[10pt]{article}
\usepackage[dvips]{graphicx}

\usepackage{epsfig}
\usepackage{graphicx}
\usepackage{indentfirst}
\usepackage{amsmath}
\usepackage{amsfonts}
\usepackage{amssymb}

\begin{document}

\begin{center}

{\Large\bf PARTICLE PRODUCTION IN AN\\[5PT]
EXPANDING UNIVERSE DOMINATED\\[5PT]
BY DARK ENERGY FLUID\\[5PT]}
\medskip

A.B. Batista\footnote{e-mail: brasil@cce.ufes.br}, J.C.
Fabris\footnote{e-mail: fabris@pq.cnpq.br. Present address:
Institut d'Astrophysique de Paris (IAP), 98bis Boulevard Arago,
75014 Paris, France.} and S.J.M. Houndjo\footnote{e-mail:
sthoundjo@yahoo.fr},
\medskip

Departamento de F\'{\i}sica, Universidade Federal do Esp\'{\i}rito
Santo\\ 29060-900, Vit\'oria, Esp\'{\i}rito Santo, Brazil
\medskip

\end{center}

\begin{abstract}

We investigate the fate of particle production in an expanding universe dominated by a perfect fluid
with equation of state $p = \alpha\rho$. The rate of particle production, using the Bogolioubov coefficients,
are determined exactly for any value of $\alpha$ in the case of a flat universe. When the strong energy condition
is satisfied, the rate of particle production decreases as time goes on, in agreement to the fact that the four-dimensional
curvature decreases with the expansion; the opposite occurs when the strong energy condition is violated. In the phantomic case, the rate of particle production
diverges in a finite time. This may lead to a backreaction effect, leading to the avoidance of the big rip singularity, specially if
$- 1 > \alpha > - \frac{5}{3}$.

\end{abstract}

Pacs numbers: 98.80.-k, 98.80.Es

\section{Introduction}

The phenomenon of particle production in an expanding universe is
due, essentially, to the fact that in curved space-time the vacuum
is not unique \cite{birrel,jacobson}. As the universe evolves, and the curvature changes,
the vacuum state also changes: the initial vacuum state,
representing the state with no particle, becomes later a multiparticle state. The particle production is directly
connected with the curvature of the universe. If the universe is
spatially flat, the cosmic evolution leads asymptotically to a
Minkowski space-time, where the phenomenon of particle production
does not occur anymore. However, this is true only if the strong
energy condition is verified: the energy density $\rho$ and the pressure
$p$ must satisfy the relation $\rho + 3p > 0$. If the strong
energy condition is violated the particle production should be zero
initially, increasing as the universe evolves, in connection with
the increase of the four-dimensional curvature with time when $\rho + 3p < 0$.
\par
In order to compute the particle production in a given
cosmological model, it is necessary to fix an initial vacuum
state. If the universe is always decelerating, there is no natural
and unique choice for this, since all modes are initially outside
the Hubble radius feeling strongly the curvature of the
space-time. However, a natural choice for the initial vacuum state
can be done for the case of an accelerating universe, since all
physical modes are initially well inside the Hubble radius, and
they behave as in a Minkowski space-time. This is one of the
reasons why the primordial inflationary scenario is so successful
theoretically: it encodes naturally a mechanism for quantum
fluctuations, which will be later the seeds for the large scale
structure of the universe \cite{jerome}.
\par
Observations indicate that we live today in an inflationary phase,
since the universe is accelerating \cite{spergel,uzan}. Hence, the
mechanism of particle production may play again a very important
role today. Suppose we define a quantum state for a given field
today. The question we want to address concerns the rate of
particle production. In particular, we would like to know if such
particle production is so important that it can alters, by
back-reaction, the later evolution of the universe. The
back-reaction due to quantum effects has already studied in the
case of the cosmological constant \cite{ford2}. The back-reaction
may be particularly relevant for the case the universe is
dominated by a phantom fluid since, classically, in such a
situation the universe will inevitably evolve towards a new
singular state, called the big rip. We remember that the
observational data favors somehow a phantom scenario today
\cite{caldwell}. A phantom fluid has many special features. Local
configurations of a phantom fluid leads to regular black holes
\cite{kirill}. In cosmology, a universe dominated by a phantom
fluid implies a future singularity that will occur in a future
finite proper time \cite{frampton}.
\par
We will solve the rate of the particle production when the
universe is flat and dominated by an equation of state $p =
\alpha\rho$, for an arbitrary value of $\alpha$. A simplified
scenario, not considering the different phase transitions that
occur in the real universe, will be adopted. The calculations will
be performed for a massless scalar field, for which the
corresponding Klein-Gordon equation is solved. Through the
quantization of the field, the rate of particle may be determined
using the technique of Bogolioubov's coefficients. The general
solution for this problem is expressed in terms of Bessel
functions. The rate of particle production is determined exactly
for any value of $\alpha$. The connection of this rate with the
strong energy condition is established. It is shown that for a
phantom fluid the rate of particle production diverges as the big
rip is approached. This may lead to a back-reaction effect
changing the effective equation of state of the universe, implying
perhaps the avoidance of the big rip itself: the particle produced
does not necessarily obeys the phantom equation of state and may
become the dominant component of the universe if $ - 1 > \alpha >
- \frac{5}{3}$. Hence, the back-reaction of the particle
production may be an ineluctable mechanism to the avoidance of the
big rip if the pressure is not excessively negative. The limiting
case, $\alpha = - \frac{5}{3}$, has already been found before in a
complete different context, that is, in the evolution of classical
scalar perturbations in phantom cosmological models
\cite{finelli,fabris}.
\par
This paper is organized as follows. In the next section we review
the formalism of particle creation in a curved space-time. In
section $3$ we apply the formalism to a perfect fluid cosmological
model, and we determine the rate of particle creation for any
value of $\alpha$. In section $4$ we present our conclusions. Even
if many steps of the formalism used and of the calculation
performed are quite well known, we will present them with some
details in order to trace the main features of the final results.

\section{Basic formalism of particle creation}

In the Minkowski spacetime, the Lorentz invariance allows to
identify a unique vacuum state. However, in a curved space-time
generally the Lorentz symmetries are not present. As a
consequence, there is no unique vacuum state, and the particle
concept (as excitation of a vacuum state) becomes ambiguous. The
nonuncity of the vacuum state is responsible for the mechanism of
"particle production", which is due, in some sense, to the fact
that different observers define different states as the "vacuum
state". The way this mechanism operates is quite model dependent.
A general way to analyse the problem is to consider a quantum
field "living"\, in a given space-time, and to obtain the particle
production via the Bogolioubov transformation that connects
different vacuum states in a given model. In the present work we
will consider a free, massless scalar field\, "living"\, in a flat
Friedmann-Robertson-Walker space-time whose dynamics is determined
by a barotropic fluid such that $p = \alpha\rho$. We will consider
$\alpha$ as an arbitrary parameter. In particular, $\alpha$ can be
negative.
\par
The Lagrangian density describing a massive scalar field $\varphi$
non-minimally coupled to gravity is given by
\begin{equation}
\label{lagrangian}
\mathcal{L}(x)=\frac{1}{2}\sqrt{-g}\left\{g^{\mu\nu}\varphi\left(x\right)_{,\,\mu}\varphi\left(x\right)_{,\,\nu}-[m^{2}+\xi\,R\left(x\right)]\varphi^{2}(x)\right\} \quad,
\end{equation}
where $R$ is the Ricci scalar and $\xi$ is a coupling constant.
The so-called minimal-coupling is obtained when $\xi = 0$, while
the conformal coupling requires $\xi = \frac{1}{6}$. We work in
natural unities such that $G=c=\hbar=1$. This Lagrangian density
implies the following equation of motion:
\begin{equation}
\label{em}
\square\varphi+m^{2}\varphi+\xi R\varphi=0
\end{equation}
The scalar product of a pair of solutions of the Klein-Gordon equation (\ref{em}) is defined by:
\begin{equation}
\label{scalar-product}
\bigg(\varphi_{1}\left(x\right),\varphi_{2}\left(x\right)\bigg)=i\int_{\Sigma}\left(\varphi_{2}^{*}(x)\stackrel{\longleftrightarrow}{\partial_{\mu}}\varphi_{1}(x)\right)[-g_{\Sigma}(x)]^{\frac{1}{2}}d\Sigma^{\,\mu}
\end{equation}
where $d\Sigma^{\,\mu}=d\Sigma \,n^{\mu}$ with $d\Sigma$ being the
volume element of the spatial-type hypersurface $\Sigma$, and
where $n^{\mu}$ is a timelike unitary vector normal to that
hypersurface. This scalar product is independent of the choice of
the hypersurface.
\par
We proceed by quantizing the scalar field in the canonical way. The conjugate momentum is defined by
\begin{equation}
\label{conjugate}
\pi=\frac{\partial \mathcal{L}}{ \partial \dot{\varphi}} \quad ,
\end{equation}
where the dot means derivation with respect to time.
The field variable and its conjugate momentum satisfy the usual commutation relations:
\begin{eqnarray}
\label{cr1}
\left[\varphi(t,x),\varphi(t,x^{\prime})\right]&=& 0 \quad ,\nonumber\\
\label{cr2}
\left[\pi(t,x),\pi(t,x^{\prime})\right]&=& 0 \quad ,\\
\label{cr3}
\left[\varphi(t,x),\pi(t,x^{\prime})\right]
&=& i \delta^{3}(x-x^{\prime})\nonumber \quad .
\end{eqnarray}
\par
Let us consider $\{\varphi_{j}\}$ the complete ensemble of positive solutions of equation (\ref{em}); so, $\{\varphi^{*}_{j}\}$
will be the complete ensemble of negative solutions of the same equation.
The field $\varphi$ may be written under the form,
\begin{equation}
\label{decomp}
\varphi = \sum_{j}(c_{j}\varphi_{j} + c^{\dagger}_{j}\varphi^{\,*}_{j})
\end{equation}
where $c_{j}$ and $c^{\dagger}_{j}$ are the annhilation and
creation operators such that
\begin{equation}
\label{cr-bis}
[c_{j},c^{\dagger}_{j}] = \delta_{jj^{\,\prime}}.
\end{equation}
The vacuum state $\left|0\right\rangle$ is defined through the
condition $c_{j}\left|0\right\rangle = 0$. In a flat space-time,
the positive solutions are given by positive frequencies, and in
this way, considering $t$ as the time coordinate, a unique vacuum
state is obtained, the Minkowski vacuum.
\par
In a curved space-time, the situation is different. In general the
choice of $\{\varphi_{j}\}$ is not unique, and as consequence the
notion of a vacuum state is not unique either. This means that it
is not possible to have a universal state describing the absence
of particle; the notion of particle itself becomes ambiguous.
\par
Let us consider an asymptotically flat space-time in the past
infinity and in the future infinity, but not in the intermediate
region. The positive frequency solutions in the past infinity are
denoted by $\varphi_{j}$, while the positive frequency solution in
the future infinity are denoted by $\phi_{j}$. These solutions
obey the following conditions:
\begin{eqnarray}
(\varphi_{j},\varphi_{j^{\,\prime}}) &=& (\phi_{j},\phi_{j^{\,\prime}})=\delta_{jj^{\,\prime}}\quad ,\\
(\varphi^{\,*}_{j},\varphi^{\,*}_{j^{\,\prime}})
&=& (\varphi^{\,*}_{j},\varphi^{\,*}_{j^{\,\prime}})=-\delta_{jj^{\,\prime}} \quad , \\
(\varphi_{j},\varphi^{\,*}_{j^{\,\prime}})
&=& (\phi_{j},\phi^{\,*}_{j^{\,\prime}})=0 \quad .
\end{eqnarray}
The field $\varphi$ may be expressed, at same time, in terms of the functions $\{\varphi_{j}\}$ or in terms of the function $\{\phi_{j}\}$ ,
since both basis are complete. Hence,
\begin{equation}
\label{description}
\varphi=\sum_{j}\left(c_{j}\varphi_{j} + c^{\dagger}_{j}\varphi^{\,*}_{j}\right)=\sum_{j}\left(b_{j}\phi_{j}+b^{\dagger}_{j}\phi^{\,*}_{j}\right) \quad .
\end{equation}
The annhilation and creation operators in the past infinity are given by $c_{j}$ and $c^{\dagger}_{j}$, while
$b_{j}$ and $b^{\dagger}_{j}$ are the corresponding operators in the future infinity. The vacuum state in the past infinity is defined by $c_{j}\left|0\right\rangle_{in}=0$  $\forall j$, describing the situation where no particle is present initially.
The vacuum state in the future infinity is defined by
$b_{j}\left|0\right\rangle_{out}=0$  $\forall \,j$ , describing a situation where no particle is present in the future infinity.
\par
The two ensemble of annhilation and creation operators are
connected by the relations
\begin{eqnarray}
c_{j} = \sum_{k}(\alpha_{kj}b_{k}+\beta_{kj}b^{\dagger}_{k}) \quad
,
\\
c^{\dagger}_{j} =
\sum_{k}(\beta_{kj}b_{k}+\alpha^{*}_{kj}b^{\dagger}_{k})
\quad , \\
b_{k} = \Sigma_{j}(\alpha^{*}_{kj}c_{j}-\beta^{*}_{kj}c^{\dagger}_{j}) \quad ,\\
b^{\dagger}_{k} =
\Sigma_{j}(\alpha_{kj}c^{\dagger}_{j} - \beta_{kj}c_{j}) \quad .
\end{eqnarray}
\par
The above relations define the Bogolioubov transformation, where
$\alpha_{jk}$ and $\beta_{jk}$ are the Bogolioubov's coefficients.
Since the asymptotic basis are complete, the in-modes can be
expressed in terms of the out-mode and vice-versa:
\begin{eqnarray}
\label{in}
\varphi_{j}=\sum_{k}\left(\alpha_{jk}\phi_{k}+\beta_{jk}\phi^{\,*}_{k}\right) \quad ,\\
\label{out}
\phi_{k}=\sum_{j}\left(\alpha^{*}_{jk}\varphi_{j}-\beta_{jk}\varphi^{*}_{j}\right) \quad .
\end{eqnarray}
In order the formalism above to be consistent, the Bogolioubov's coefficients must obey the following relations:
\begin{eqnarray}
\sum_{j}\left(\alpha^{*}_{jk}\alpha_{jj^{\prime}}-\beta_{jk}\beta^{*}_{jj^{\prime}}\right) &=& \delta_{kj^{\prime}} \quad ,\\
\sum_{j}\left(\alpha^{*}_{jk}\beta_{jj^{\prime}}-\beta_{jk}\alpha^{*}_{jj^{\prime}}\right) &=& 0 \quad .
\end{eqnarray}
\par
Now, we can describe the phenomenon of particle creation by a time-dependent gravitational field. Using the Heisenberg representation, $\left|0\right\rangle_{in}$ is the state system at each moment. The number operator counting the number of particles in the future infinity is  $N_{k}=b^{\dagger}_{k}b_{k}$.
So, the number of particles in the future infinity is given by:
\begin{eqnarray}
\left\langle N_{k}\right\rangle &=&  \left\langle0,in\left|b^{\dagger}_{k}b_{k}\right|0,in\right\rangle\nonumber\\
&=& \sum_{j}\left|\beta_{kj}\right|^{2}
\end{eqnarray}
A non null $\beta_{jk}$ implies particle creation.

\section{Particle creation in perfect fluid cosmological models}

The Robertson-Walker metric describing a four dimensional
space-time flat, written in terms of the conformal time, is:
\begin{equation}
\label{metricbis}
ds^{2}=a^{2}(\eta)\left[\,d\eta^{2} - dx^{2} - dy^{2} - dz^{2}\,\right] \quad .
\end{equation}
\par
In terms of the conformal time, the Einstein's equations take the following form:
\begin{eqnarray}
\label{ee1}
\frac{\left(a^{\prime}\right)^{2}}{a^{2}}=\frac{8\pi G}{3}\rho \,a^{2} \quad ,\\
\label{ee2}
2\frac{a^{\prime\prime}}{a}-\frac{\left(a^{\prime}\right)^{2}}{a^{2}}=-8\pi G p\, a^{2}
\end{eqnarray}
where $\rho$ and $p$ are the energy density and the pressure, respectively. We suppose that they are
related by a barotropic equation of state such that
\begin{equation}
\label{eos}
p = \alpha\rho \quad .
\end{equation}
The fact that the sound velocity in this fluid must be equal or smaller than the velocity of the light implies that $\alpha \leq 1$.
From the equations of motion (\ref{ee1},\ref{ee2}), we obtain
\begin{equation}\label{sf}
a(\eta)=a_{0}\,|\eta|^\beta \quad , \quad \beta = \frac{2}{1 +
3\alpha} \quad .
\end{equation}
If $\alpha > - 1/3$, the times evolves such that $0 < \eta <
\infty$, while, for $\alpha < - 1/3$, the interval is $- \infty <
\eta < 0_-$ This class of cosmological solutions does not
interpolate two minkowskian space-time: the four dimensional
curvature diverges for $\eta \rightarrow 0$ and goes to zero for
$\eta \rightarrow \infty$. In the region the curvature goes to
zero the modes are well inside the horizon and behave as in the
Minkowski space-time. This vacuum is called Bunch-Davies vacuum
and it plays a fundamental role in the origin of quantum
fluctuations in the inflationary scenarios for structure
formation. Fixed the initial vacuum state, the formalism described
in the preceding section applies for the flat FRW cosmological
models and the temporal evolution of the occupation number
operator can be computed.
\par
Let us write the solutions of the Klein-Gordon equation under the form
\begin{equation}
\label{fourier}
\varphi\left(\eta,\vec{x}\right) = \int dk^{3}\left[c_{\vec{k}}\varphi_{k}\left(\eta,\vec{x}\right)+ c^{\dagger}_{\vec{k}}\varphi^{*}_{k}\left(\eta,\vec{x}\right)\right] \quad ,
\end{equation}
where the functions $\varphi_{k}\left(\eta,\vec{x}\right)$, through a separation of variable, have the form
\begin{equation}
\label{redefinition}
\varphi_{k}\left(\eta,\vec{x}\right)=\left(2\pi\right)^{-\frac{3}{2}}e^{i\,\vec
k\cdot\vec x}\,\frac{\chi_{k}(\eta)}{a(\eta)}
\end{equation}
\par
Using the flat Robertson-Walker metric, the massless Klein-Gordon equation takes the form,
\begin{equation}
\label{em1}
\varphi^{\prime\prime}_{k} + 2\frac{a^{\prime}}{a}\varphi_{k}+k^{2}\varphi_{k} = 0 \quad .
\end{equation}
Inserting (\ref{redefinition}) into (\ref{em1}), we obtain
\begin{equation}
\label{parametric}
\chi_{k}\,^{\prime\prime}(\eta)+\left(k^{2}-\frac{a^{\prime\prime}}{a}\right)\chi_{k}(\eta) = 0 \quad .
\end{equation}
Using the expression for the scale factor, the above equation reduces to
\begin{equation}
\label{parametricbis}
\chi_{k}\,^{\prime\prime}(\eta)+\left(k^{2}-\frac{\beta(\beta-1)}{\eta^2}\right)\chi_{k}(\eta)
= 0 \quad .
\end{equation}
This is essentially the same equation governing the evolution of gravitational waves in an expanding universe \cite{grishchuk}.
\par
This equation admits solutions under the form of Hankel's functions:
\begin{equation}
\label{50} \chi_{k}(\eta) =
\eta^\frac{1}{2}\biggr[A_{k}H^{(1)}_{q}(k|\eta|) +
B_{k}H^{(2)}_{q}(k|\eta|)\biggl] \quad , \quad q = |\frac{1}{2} -
\beta| \quad ,
\end{equation}
$H^{(1,2)}_{k}$ are the Hankel's functions, ($A_{k}$, $B_{k}$) are
constants whose values are fixed by normalizing the corresponding
modes. To do this, we use the Wronskian relation
\begin{equation}
zH^{(2)}_{q}(z)\partial_{z}H^{(1)}_{q}(z)-zH^{(1)}_{q}(z)\partial_{z}H^{(2)}_{q}(z)
= \frac{4i}{\pi} \quad .
\end{equation}
Imposing the orthonormalisation of the modes one obtains:
\begin{equation}
\left|A_k\right|^2 = \left|B_{k}\right|^{2} = \frac{\pi}{4} \quad .
\end{equation}
The function $\chi_k(\eta)$ becomes (up to an arbitrary phase)
\begin{equation}
\label{sol}
\chi_{k}(\eta)=\frac{\sqrt{\pi
\eta}}{2}\biggr(\epsilon_1H^{(1)}(k|\eta|) +
\epsilon_2H^{(2)}_{q}(k|\eta|)\biggl) \quad , \quad \epsilon_{1,2}
= 0,1 \quad .
\end{equation}
This solution has been already studied previously (see
\cite{ford1}), but in view of a primordial cosmological models, in
special in de Sitter or quasi-de Sitter phase.
\par
With the aim of determining the particle production for the cosmological model, we will rewrite the Bogolioubov's transformations in a convenient way.
For a minimal coupling and a massless field, the Lagrangian density takes the form:
\begin{equation}
\label{lagran}
\mathcal{L}(x) = \frac{1}{2}\left[-g\left(x\right)\right]^{\frac{1}{2}}\left\{g^{\mu\nu}\varphi\left(x\right)_{,\,\mu}\varphi\left(x\right)_{,\,\nu}\right\} \quad .
\end{equation}
With the background described above, the Lagrangian can be rewritten as
\begin{equation}
\label{lagranbis}
\mathcal{L} = \frac{1}{2}\left\{\chi^{\,\prime}\,^{2}-k^{2}\chi^{2}-2\frac{a^{\prime}}{a}\chi\chi^{\prime}+\left(\frac{a^{\prime}}{a}\right)^{2}\chi^{2}\right\}
\quad .
\end{equation}
The conjugate momentum is
\begin{equation}
\label{60}
p = \frac{\partial L}{\partial \chi'} \quad .
\end{equation}
Hence, the Hamiltonian is
\begin{equation}
\label{hamil}
\mathcal{H} = \frac{1}{2}\left\{p^{2} + k^{2}\chi^{2} + 2\frac{a^{\prime}}{a}\chi p\right\} \quad .
\end{equation}
After quantization, the operators $\mu$ and $p$ satisfy the commutation relation
\begin{eqnarray}
\label{cr1}
\left[\chi ,p\right] = i \quad .
\end{eqnarray}
We can now determine the expressions for the creation and annhilation operators,
$b^{\dagger}$ and $b$ respectively, with respect to the Hamiltonian (\ref{hamil}).
To do this, we fix:
\begin{equation}
\chi = \frac{1}{\sqrt{2k}}(b + b^{\dagger}) \quad , \quad p = - i\sqrt{\frac{k}{2}}(b - b^{\dagger}) \quad .
\end{equation}
These operators satisfy the commutation relation
\begin{equation}
\label{crbis}
\left[b,b^{\dagger}\right] = 1 \quad .
\end{equation}
The Hamiltonian takes the form
\begin{equation}
\mathcal{H} =k b^{\dagger}b + \sigma(\eta){b^{\dagger}}^{2}+\sigma^{*}(\eta)b^{2}+ \frac{i}{2}(k+\frac{a^{\prime}}{a}) \quad,
\end{equation}
where the coupling function is given by $\sigma(\eta)=ia^{\prime}/2a$.
Using the normal ordering
\begin{equation}
:\mathcal{H}:=\mathcal{H}-\left\langle0\left| \mathcal{H}\right|0 \right\rangle \quad ,
\end{equation}
one obtains
\begin{equation}
\label{hamilbis}
\mathcal{H} = k b^{\dagger}b + \sigma(\eta){b^{\dagger}}^{2}+\sigma^{*}(\eta)b^{2} \quad .
\end{equation}
\par
We will use the Heisenberg representation for which the operators
evolve with time and the quantum states remain fix. Since there
are two independent modes, the Hamiltonian (\ref{hamilbis}) can be
written as a sum of two Hamiltonians $\mathcal H_{1}$ and
$\mathcal H_{2}$ with the same frequency:
\begin{eqnarray}
\label{hamil1}
\mathcal H_{1} = k b_{1}^{\dagger}b_{1} + \sigma(\eta){b_{1}^{\dagger}}^{2}+\sigma^{*}(\eta)b_{1}^{2} \quad ,\\
\label{hamil2}
\mathcal H_{2} = k b_{2}^{\dagger}b_{2} + \sigma(\eta){b_{2}^{\dagger}}^{2}+\sigma^{*}(\eta)b_{2}^{2} \quad .
\end{eqnarray}
Thus, the Hamiltonian becomes
\begin{equation}
\mathcal H = k\left[ b_{1}^{\dagger}b_{1}+b_{2}^{\dagger}b_{2}\right] + \sigma(\eta)\left[{b_{1}^{\dagger}}^{2}+{b_{2}^{\dagger}}^{2}\right]+\sigma^{*}(\eta)\left[b_{1}^{2}+b_{2}^{2}\right] \quad .
\end{equation}
\par
We now define two annhilation operators with momenta $\vec k$ and $- \vec k$:
\begin{align}\label{85}
c_{\vec{k}}=\frac{b_{1}-ib_{2}}{\sqrt{2}} \quad , \quad c_{-\vec{k}}=\frac{b_{1}+ib_{2}}{\sqrt{2}}\quad .
\end{align}
The Hamiltonian may be now rewritten in terms of these new operators:
\begin{equation}
\label{hamiltertio}
\mathcal H = kc_{\vec{k}}^{\dagger}c_{\vec{k}}+ kc^{\dagger}_{-\vec{k}}c_{-\vec{k}}+2\sigma(\eta)c^{\dagger}_{\vec{k}}c^{\dagger}_{-\vec{k}}+2\sigma^{*}(\eta)c_{\vec{k}}c_{-\vec{k}} \quad .
\end{equation}
The new operators obey the following equations:
\begin{eqnarray}
\frac{dc_{\vec{k}}}{d\eta} &=& - i\left[c_{\vec{k}},H\right] \quad ,\\
\frac{dc^{\dagger}_{\vec{k}}}{d\eta} &=& - i\left[c^{\dagger}_{\vec{k}},H\right] \quad .
\end{eqnarray}
Using (\ref{hamiltertio}), we find:
\begin{eqnarray}
\frac{dc_{\vec{k}}}{d\eta} = - ikc_{\vec{k}}+\frac{a^{\prime}}{a}\,c^{\dagger}_{-\vec{k}} \quad ,
\frac{dc^{\dagger}_{\vec{k}}}{d\eta}=ikc^{\dagger}_{\vec{k}}+\frac{a^{\prime}}{a}\,c_{-\vec{k}} \quad ,
\end{eqnarray}
whose solutions can be written as
\begin{eqnarray}
\label{a1}
c_{\vec{k}}(\eta)=u_{k}(\eta)c_{\vec{k}}(0) + v_{k}(\eta)c^{\dagger}_{-\vec{k}}(0) \quad , \\
\label{a2}
c^{\dagger}_{\vec{k}}(\eta)=u^{*}_{k}(\eta)c^{\dagger}_{\vec{k}}(0)+v^{*}_{k}(\eta)c_{-\vec{k}}(0) \quad ,
\end{eqnarray}
where $c_{\vec{k}}(0)$  and  $c^{\dagger}_{\vec{k}}(0)$ are the
initial values of the operators $c_{\vec{k}}(\eta)$ and
$c^{\dagger}_{\vec{k}}(\eta)$, respectively. The function $v_k$ is
nothing else than the Bogolioubov coefficient $\beta_{ij}$ defined
before.
\par
From (\ref{a1}), evaluate at $\eta=0$, one obtains:
\begin{equation}
c_{\vec{k}}(0)=u_{k}(0)c_{\vec{k}}(0)+v_{k}(0)c^{\dagger}_{-\vec{k}}(0) \quad ,
\end{equation}
resulting that $u_{k}(0) = 1$ and $v_{k}(0) = 0$.
Due to the fact that $\left[c_{\vec{k}}(\eta),c^{\dagger}_{\vec{k}}(\eta)\right]=1$, one can write
\begin{equation}
\left[u_{k}(\eta)c_{\vec{k}}(0)+v_{k}(\eta)c^{\dagger}_{-\vec{k}}(0)
\quad , \quad
u^{*}_{k}(\eta)c^{\dagger}_{\vec{k}}(0)+v^{*}_{k}(\eta)c_{-\vec{k}}(0)\right]
= 1 \quad ,
\end{equation}
resulting that
\begin{equation}
\left|u_{k}(\eta)\right|^{2}-\left|v_{k}(\eta)\right|^{2} = 1 \quad .
\end{equation}
That is the normalization condition. We find also the relations
\begin{eqnarray}
\frac{du_{k}(\eta)}{d\eta}=-ik\, u_{k}(\eta)+\, \frac{a^{\prime}}{a}\, v^{*}_{k}(\eta) \quad , \\
\frac{dv_{k}(\eta)}{d\eta}=-ik\, v_{k}(\eta)+\, \frac{a^{\prime}}{a}\, u^{*}_{k}(\eta)
\end{eqnarray}
The sum ($u_{k}+v^{*}_{k}$) satisfies the relation
\begin{equation}
\left(u_{k}+v^{*}_{k}\right)^{\prime\prime}+\left(k^{2}-\frac{a^{\prime\prime}}{a}\right)\left(u_{k}+v^{*}_{k}\right)=0 \quad .
\end{equation}
Moreover,
\begin{equation}
\chi_{k}(\eta) = u_{k}(\eta) + v^{*}_{k}(\eta) \quad ,
\end{equation}
leading to
\begin{equation}
v^{*}_{k}(\eta)=\left[\frac{1}{2}+i\frac{\beta}{2k\eta}\right]\chi_{k}(\eta)- \frac{i}{2k}\chi^{\,\prime}_{k}(\eta) \quad .
\end{equation}
\par
Using the solutions (\ref{sol}) for the field $\chi_k(\eta)$, with
$\epsilon_2 = 0$, we find
\begin{equation}
v^{*}_{k}(\eta) = \frac{\sqrt{\pi
|\eta|}}{4}e^{i\theta_{k}}\left[H^{(1)}_{q}(k|\eta|)-iH^{(1)}_{q-1}(k|\eta|)\right]
\quad .
\end{equation}
Hence,
\begin{eqnarray}
\left|v_{k}(\eta)\right|^{2}=\frac{\pi|\eta|}{16}\Bigg[H^{(1)}_{q}(k|\eta|)H^{(2)}_{q}(k|\eta|)+H^{(1)}_{q-1}(k|\eta|)H^{(2)}_{q-1}(k|\eta|)\nonumber\\
+i\left(H^{(1)}_{q}(k|\eta|)H^{(2)}_{q-1}(k|\eta|)-H^{(1)}_{q-1}(k|\eta|)H^{(2)}_{q}(k|\eta|)\right)\Bigg]
\quad .
\end{eqnarray}
This expression gives the rate of particle production. It has the following asymptotic behaviours:
\begin{eqnarray}
k|\eta| \rightarrow 0 \quad , \quad q < \frac{1}{2} \quad (\alpha
> \frac{1}{3}) \quad &\Rightarrow& \quad
|v_k(\eta)|^2 \rightarrow \frac{\pi}{8k}\biggr(\frac{k|\eta|}{2}\biggl)^{3q - 2} \quad ;\\
k|\eta| \rightarrow 0 \quad , \quad q = \frac{1}{2} \quad (\alpha = \frac{1}{3}) \quad &\Rightarrow& \quad |v_k(\eta)|^2 \rightarrow \frac{\pi}{4k} \quad ;\\
k|\eta| \rightarrow 0 \quad , \quad q > \frac{1}{2} \quad (\alpha
< \frac{1}{3}) \quad &\Rightarrow& \quad
|v_k(\eta)|^2 \rightarrow \frac{\pi}{8k}\biggr(\frac{k|\eta|}{2}\biggl)^{1-2q} \quad ;\\
k|\eta| \rightarrow \infty \quad , \quad \forall q \quad
&\Rightarrow& \quad |v_k(\eta)|^2 \rightarrow 0\quad .
\end{eqnarray}
\par
For $\alpha > - 1/3$, $\eta \rightarrow 0$ implies $t \rightarrow
0$, while for $\alpha < - 1/3$, $\eta \rightarrow 0$ implies $t
\rightarrow \infty$. The asymptotic behavior above just show that,
for $\alpha > -1/3$, there is no natural initial vacuum state,
except for the radiative case $\alpha = 1/3$: the initial state is
fixed arbitrarily by hand. That is, an decelerating universe
requires a previous accelerated phase in order to establishes a
natural vacuum initial state. For $\alpha < - 1/3$ there is, in
opposition, a natural initial vacuum state, and the particle
production is initially zero, becoming infinite asymptotically.
\par
In general, the Hubble radius is given by $d_H(\eta) \propto
\eta^{-\frac{3(1 + \alpha)}{1 + 3\alpha}}$. The Hubble radius
increase with time for $\alpha > - 1$, remaining constant for
$\alpha = - 1$, and it decreases for $\alpha < - 1$. Hence, in the
phantom case, all modes are well inside the Hubble radius if we go
back far enough in the past, and it is quite reasonable to assume
that the initial state is the vacuum state \footnote{In the real
universe we should perhaps prefer to consider an initial non
vacuum state, since there was already a particle production in the
previous phases of the universe. But, an initial non vacuum state
does not change our main conclusions, except for some very special
cases.}. In figure $1$ the behavior of the scale factor, of the
Hubble radius and of an arbitrary physical model are displayed for
a phantom cosmological model, illustrating the situation we have
just described.

\begin{figure}[!t]
\includegraphics[width=\linewidth]{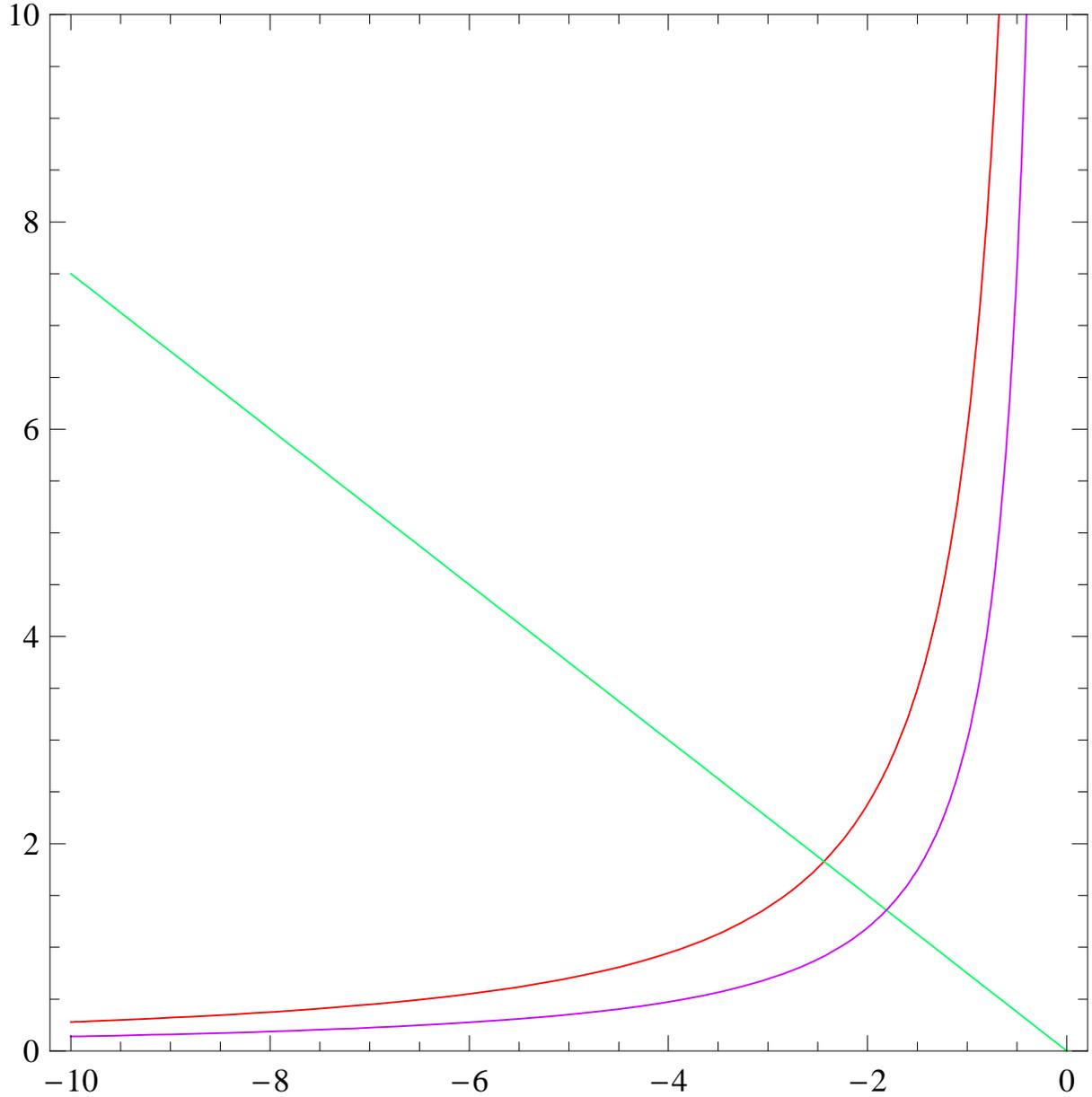}
\caption{{\protect\footnotesize Behavior, as function of time, of
the scale factor (red), of the Hubble radius (green) and of a
typical physical mode (violet) in a phantom model.}}
\end{figure}

\par
For the phantom scenario the future asymptotic occurs in a finite
proper time. Hence, the particle production may change the final
state; in particular, it may lead to the avoidance of the big rip.
This may happens since the equation of state of the particles
created, for a massless scalar field, is given by $\rho_s = p_s$,
and we can expect that the phantom fluid will not become the
dominant component anymore. In fact, the energy density of perfect
fluid scales as $\rho_{ph} \propto a^{-3(1 + \alpha)} \propto
\eta^{-\frac{6(1 + \alpha)}{1 + 3\alpha}}$. If we suppose the
energy density of the scalar particle created scales according to
the rate evaluated above, we have $\rho_s \propto \eta^\frac{4}{1
+ 3\alpha}$. So the ratio $\rho_{ph}/\rho_s$ goes to zero if $- 1
> \alpha > - 5/3$ and to infinity if $\alpha < - 5/3$: the
back-reaction can be effective if the equation of state of the
phantom fluid is not deeply negative.
\par
The particular case $\alpha = - \frac{1}{3}$  must be studied separately.
In terms of the conformal time, the solution for this case takes the form
\begin{equation}
a(\eta)=a_{0}\,e\,^{\lambda\eta} \quad .
\end{equation}
where $a_{0}$ and $\lambda$ are constants.
The Klein-Gordon equation now reads,
\begin{equation}
\chi_{k}\,^{\prime\prime}(\eta)+D\,^{2}\chi_{k}(\eta) = 0 \quad ,
\end{equation}
with
\begin{equation}
D\,^{2}=k\,^{2}-\lambda^{2} \quad .
\end{equation}
\par
There is no propagation of the quantum modes if $D^2 \leq 0$. However, for $D\,^{2}>0$, the solution becomes
\begin{equation}
\chi_{k}(\eta) = \chi_{0}e^{-iD\eta} \quad .
\end{equation}
The normalisation implies that
\begin{equation}
\chi_{0}=\left(2D\right)^{-\frac{1}{2}} \quad .
\end{equation}
\par
Using the last expression, we obtain
\begin{equation}
v^{*}_{k}(\eta)  = \frac{1}{2}\frac{1}{\sqrt{2D}}\biggr[1 - \frac{D}{k} + i\frac{\lambda}{k}\biggl] \quad .
\end{equation}
Particles, for this case, are created with a constant rate during
all the evolution of the universe for the modes that are inside
the Hubble radius, while no particle is created for the modes that
are outside the Hubble radius. Note that, in this case, a mode
that is initially inside the Hubble radius remains always inside
it.

\section{Conclusions}

We have evaluate the rate of creation of massless scalar particle
in a universe dominated by a perfect fluid whose equation of state
is given by $p = \alpha\rho$. An analytic expression has been
found in terms of Hankel's functions. Since we have not considered
a complete cosmological scenario, with a sequence of different
phases, with an initial inflationary phase (as in standard
cosmological model), the calculation performed here makes sense,
strictly speaking, only when the strong energy condition is
violated, that is, $\alpha < - 1/3$. For such a case, there is a
natural initial vacuum state, from which the particle occupation
number can be determined. However, we formally extended the
calculation for any value of $\alpha$. It has been found that for
those "dark energy" scenarios the rate of particle creation
diverges as $t \rightarrow \infty$, that is, in the future
infinity.
\par
We have payed special attention to the phantom scenario. The
reason is that a universe dominated by a phantom fluid may develop
a singularity in a finite future time, the so-called big rip. In
this case, we have found that it is possible that the energy
density associated to the particles created (which obey in present
case an equation of state of the type $p_s = \rho_s$) becomes
dominant over the phantom fluid if $- 1 > \alpha > - 5/3$. Hence,
the big rip can be avoided if the pressure is not deeply negative.
\par
It is interesting to remark that a similar critical point, $\alpha
= - 5/3$, has been found in the case of classical scalar
perturbation \cite{fabris}. In that case, however, the evolution
of scalar perturbation may destroy the conditions of homogeneity
(necessary for the big rip) if $\alpha < - 5/3$, that is, if the
pressure is negative enough. The result found here is exactly the
opposite: quantum effects can be operative in the sense of
destroying the conditions for the big rip if the pressure is not
negative enough. It must be stressed, however, that the evaluation
made in the present work must be complemented by a study of more
general quantum fields and by a deeper thermodynamical analysis of
the energy balance between the phantom fluid and the created
particles as the big rip is approached.
\newline
 \vspace{0.5cm}
 \newline
 {\bf Acknowledgements}:\\
 We thank CNPq (Brazil) and the french/brazilian
scientific cooperation CAPES/COFECUB (project number 506/05) for
partial financial support. We thank specially J\'er\^ome Martin
for his criticisms on the text.

\end{document}